\documentclass[fleqn,twoside]{article}
\usepackage{espcrc2}

\usepackage{graphicx}
\usepackage[figuresright]{rotating}

\newcommand{\AmS}{{\protect\the\textfont2
  A\kern-.1667em\lower.5ex\hbox{M}\kern-.125emS}}

\title{ Phenomenological Models of the
Quark-Gluon Plasma Equation of State}

\author{Peter N. Meisinger 
	\address[stupidaddressmark] {Dept. of Physics, 
        Washington University, \\ 
        St. Louis, MO 63130 USA}%
        \thanks{We gratefully acknowledge the support of the U.S. Dept. of
		Energy under DOE DE-FG02-91ER40628},
        Travis R. Miller \addressmark,
        Michael C. Ogilvie \addressmark[stupidaddressmark]}
       
\begin{document}

\begin{abstract}
Two phenomenological models describing an $SU(N)$ quark-gluon
plasma are presented. The first is obtained from high temperature expansions
of the free energy of a massive gluon, while the second is derived by
demanding color neutrality over a certain length scale. Each model has a
single free parameter, exhibits behavior similar to lattice simulations over
the range $T_{d}-5T_{d}$, and has the correct blackbody behavior for large
temperatures. The $N=2$ deconfinement transition is second order in both
models, while $N=3$,$4$, and $5$ are first order. Both models
appear to have a smooth large-$N\,\ $limit.
In both models, the confined
phase is characterized by a mutual repulsion of Polyakov loop
eigenvalues that makes the Polyakov loop expectation value zero. In the
deconfined phase, the rotation of the eigenvalues in the complex plane
towards $1$ is responsible for the approach to the blackbody limit over the
range $T_{d}-5T_{d}$.
The addition of quarks in $SU(3)$ breaks $%
Z(3)\,$symmetry weakly and eliminates the
deconfining phase transition for 
sufficiently light quarks.

\end{abstract}

\maketitle

\section{Introduction}
\label{sec:intro}

We have developed\cite{Meisinger:2001cq}
two simple models for
the $SU(N)$ gluon plasma equation of state,
both based on the use of 
the Polyakov loop $P$ as the natural order parameter of the
deconfinement transition in pure gauge theories\cite{Yaffe:1982qf}.
In these models, the eigenvalues of the
Polyakov loop are the essential degrees of freedom rather than
the Polyakov loop trace alone,
a possibility also recently explored by Pisarski
\cite{Pisarski:2000eq}.

Since $P$ is unitary,
we can parametrize its diagonal form
by $N$ phases $\theta_j$ which we refer to as eigenvalues.
In our models, 
confinement is obtained from a set of spatially constant
eigenvalues which
make $\left\langle Tr\,P\right\rangle =0$.
This is naturally obtained from a uniform distribution of
eigenvalues around the unit circle, constrained by the unitary of $P$.
In a pure gauge theory below the deconfinement temperature $T_{d}$%
, the eigenvalues are frozen in this uniform distribution. 
As $T$ moves upward from $T_d$,
the eigenvalues of the Polyakov loop rotate towards $\theta =0$ or
one of its $Z(N)\,$equivalents, and $P$ moves towards an element of the
center. In the case of a first order transition, the eigenvalues jump at $%
T_{d}$.
This motion of the eigenvalues is
responsible for the approach to the blackbody limit over the range $%
T_{d}-5T_{d}$.

As a consequence of asymptotic freedom, the perturbative
expression for the free energy as a function of the Polyakov loop
eigenvalues will be valid at sufficiently high temperatures.
At one-loop order, the result has the form
$f_{pert}=T^{4}F_4(\theta_j )$
where $F_4$ has a simple expression in terms of the fourth Bernoulli
polynomial;
$f_{pert}$
gives the blackbody behavior expected at high temperature. 
Note that $f_{pert}$ is not a
function of $\left\langle Tr\,P\right\rangle$,
instead depending directly on the eigenvalues.
This distinction begins to matter with $SU(4)$, because
for $ N \geq 4$, the eigenvalues cannot be determined solely
from  $\left\langle Tr\,P\right\rangle$.

\section{Models A and B}
\label{sec:modelab}

The first of our two models is obtained by adding, by hand, a mass $M$
to the gauge bosons, and working with the high temperature expansion of the
resultant free energy to order $M^2T^2$.
The result is
\begin{equation}
f_{A}\left( \theta \right) =
T^{4}F_4(\theta_j )- M^2 T^2 F_2(\theta_j )
\end{equation}
where $F_2$ is given as sums of second Bernoulli polynomials
\cite{Meisinger:2001fi}.
We stress that
this derivation merely indicates
the type of additional terms that
might appear in the free energy. 
The Bernoulli
polynomials appear naturally as class functions which are almost
everywhere polynomials in the $\theta$'s,
and are well-suited for the construction of a Landau
theory in the eigenvalues. 

Our second model is obtained by supposing
that there is a natural scale $R$ in position space over
which color neutrality is enforced. In other words, net color is allowed in
volumes of less than $R^{3}$, but the net color on larger scales is zero.
This leads to 
\begin{equation}
f_{B}\left( \theta \right) =f_{pert}\left( \theta \right) -\frac{1}{\beta
R^{3}}\ln \left[ J\left( \theta \right) \right] 
\end{equation}
where $J\left( \theta \right) $ is the Jacobian associated with Haar
measure on $SU(N)$.

In both models, the $T^{4}$ term, which favors the deconfined
phase, dominates for large $T$.
The other term,
which favors the confined phase, dominates for small T.
The deconfinement phase transition results from a
conflict between the two terms. 

\section{$SU(N)$ \protect\vspace{1pt}Thermodynamics for $N=2-5$%
}
\label{sec:thermo}

In the case of $SU(2)$ and $SU(3)$, model A
is analytically tractable.
The deconfining transition is second order
for $SU(2)$ and first order for $SU(3)$,
in accord with universality arguments 
\cite{Yaffe:1982qf}.
Numerical analysis for the cases of $SU(4)$ and $SU(5)$
show that the deconfinement transition is first order;
recent lattice results for $SU(4)$ also indicate a
first order transition\cite{Wingate:2001bb}.
Identical results on the order of the transition are
obtained numerically for model B.

In the case of $SU(3)$, both models show good agreement
with lattice simulations\cite{Boyd:1996bx}
upon comparing the pressure $p$,
the energy density $\varepsilon $,
and the dimensionless interaction measure
$\Delta =(\varepsilon -3p)/T^{4}$
as a function of the dimensionless variable $T/T_{d}$.
The range of temperatures over which
metastable behavior occurs can be easily determined,
and confirms the weakly first order character of the
transition.

\begin{figure}
\vspace{-0.3in}
\includegraphics[width=3in]{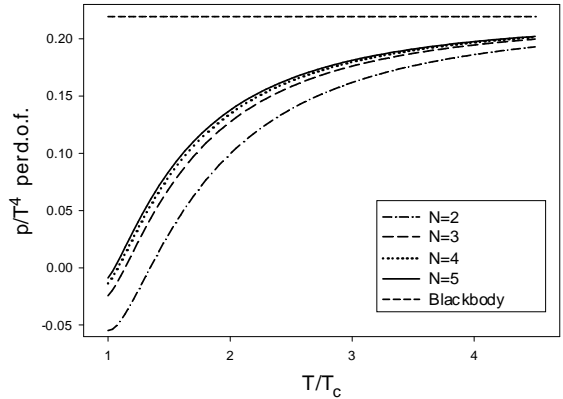}
\vspace{-0.6in}
\caption{$p/T^4$ versus $T/T_c$ for model A.}
\label{b2-p}
\vspace{-0.27in}
\end{figure}
\begin{figure}
\includegraphics[width=3in]{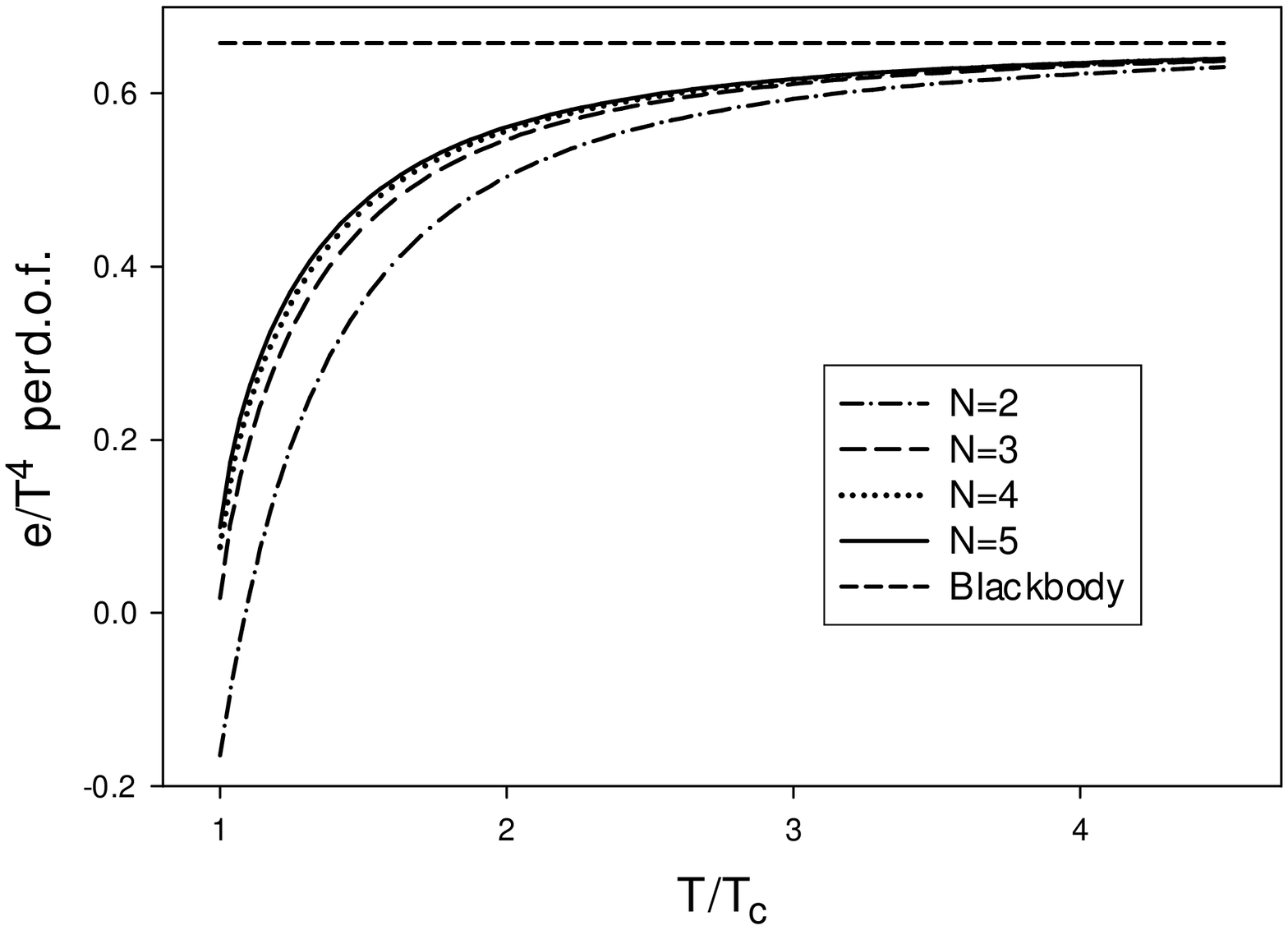}
\vspace{-0.6in}
\caption{$e/T^4$ versus $T/T_c$ for model A.}
\label{b2-e}
\vspace{-0.27in}
\end{figure}
\begin{figure}
\includegraphics[width=3in]{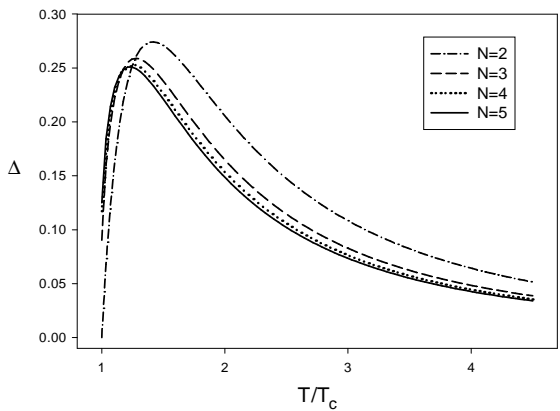}
\vspace{-0.6in}
\caption{$\Delta$ versus $T/T_c$ for model A.}
\label{b2-d}
\vspace{-0.27in}
\end{figure}  

It is enlightening to plot $p$, $\varepsilon $, and $\Delta $, each divided
by $N^{2}-1$, versus $T/T_{d}$ for $N=2$, $3$, $4$, and $5$. Figures 1-3
show the results for model A;
model B is similar.
Both models appear to 
quickly approach a large-$N$ limit.
Both models show power law
behavior in $\Delta $ for sufficiently large $T$, consistent with $\Delta
\propto 1/T^{2}$. This is the asymptotic behavior found analytically for $%
SU(2)$ in model A, and is compatible with the 
$SU(3)$ lattice data.
In camparison,
the Bag model\cite{Cleymans:1986wb},
predicts a $1/T^{4}$ behavior for $\Delta $,
which is ruled out by lattice results.

\section{Quarks}
\label{sec:quarks}

Quark effects can be included
as the free energy of quarks propagating in a
constant Polyakov loop background.
We limit
ourselves here to a discussion of very heavy quarks 
\cite{Meisinger:1995qr}
and the leading order
effect of light quarks, 
which do not depend on chiral symmetry effects
\cite{Gross:1981br,Weiss:1981rj,Weiss:1982ev}.
The expected effect of very heavy quarks is to lower the critical temperature
\cite{Green:1984sd,Ogilvie:1984ss}.
This line of first-order critical points in the $T-m$
plane terminates in a second-order end point at some finite
value of $m$.
For light quarks, we expect that the deconfinement
transition is replaced
by a smooth crossover with a rapid rise in all thermodynamic quantities.
We have confirmed both sets of behaviors in the case of $SU(3)$.

\section{Conclusions}
\label{sec:conclusion}

We have developed two simple
phenomenological equations of state for the
quark-gluon plasma, which reproduce much of the thermodynamic behavior seen
in lattice simulations.
Both models 
correctly predict the order of the deconfining phase transitions for $SU(2)$
(second order) and $SU(3)$ (weakly first order), and predict first
order transitions in $SU(4)$ and $SU(5)$ as well. 
Both models appear to have smooth large-$N$ limits.
The numerical value of the parameters introduced are reasonable for 
both models in the case of $%
SU(3)$. 
The deconfinement temperature in a pure $SU(3)$ gauge
theory is about $270\,MeV$, giving a value for $M$ of $596\,MeV$,
a plausible value for a constituent gluon mass. In model
B, we find that $R\,$is $1\,$fermi.
It is clear that by allowing the parameters $M$ and $%
R $ to depend on the temperature, a better fit to lattice data can be
obtained at the cost of introducing additional phenomenological parameters.

The success of these phenomenological models suggests a
point of view on the nature of the deconfinement mechanism.
In confining theories,
the eigenvalue distributions of the Polyakov loop are peaked
at low temperature around values evenly spaced about the unit circle in
such a way that the expectation value of the Polyakov loop is zero.
At the deconfining
transition temperature, the peaks of the eigenvalue distributions
move towards $1$, which is the asymptotic limit as $T$ goes to infinity.
We believe that this picture of deconfinement coupled with a field-theory
inspired model of chiral symmetry breaking has the potential to fully model
the equation of state of the quark-gluon plasma
\cite{Meisinger:1996kp,Dumitru:2001in}.

\end{document}